\documentclass[12pt]{emulateapj}
\usepackage{graphics}

\begin{document}

\shorttitle{$i'$-band drop-out galaxies at $z\approx 6$ in the UDF}
\title{A Matched Catalogue of $i'$-band drop-out galaxies at $z\approx 6$ in the Ultra Deep Field}
\shortauthors{Bunker \& Stanway}
\author{Andrew J.\ Bunker}
\affil{School of Physics, University of Exeter, Stocker Road, Exeter, EX4 4QL, U.K.\\ {\tt email:
bunker@astro.ex.ac.uk}}
\author{Elizabeth R.\ Stanway}
\affil{Institute of Astronomy, University of Cambridge,
Madingley Road, Cambridge, CB3\,0HA, U.K.\\ {\tt email:
ers@ast.cam.ac.uk}}

\begin{abstract}
  A recent preprint by Yan \& Windhorst (astro-ph/0407493)
  independently repeats our selection of candidate $z\approx 6$
  galaxies in the Hubble Ultra Deep Field (astro-ph/0403223). The
  agreement of this independent study with our original work is
  excellent, and most of the $i'$-band drop-out galaxies are reproduced.  To
  avoid confusion with various ID numbers for sources, we present the
  community with a matched list of $i'$-band drop-outs in the Hubble
  Ultra Deep Field.
 \end{abstract}


\maketitle

 
In March 2004 we used the publically-released imaging of the Hubble
Ultra Deep Field (UDF) to identify candidate $z\approx 6$ galaxies
through their large $(i'-z')$ colour, due to the Lyman-$\alpha$ break
at high redshift. In astro-ph/0403223 (Bunker et al.\ 2004) we
identified 54 candidate galaxies with $(i'-z')_{AB}> 1.3$ and
$z'_{AB}< 28.5$ ($8\,\sigma$ detection), corresponding to unobscured
star formation rates as low as $1\,M_{\odot}\,{\rm yr}^{-1}$ at
$z\approx 6$ (equivalent to $0.1\,L^{*}_{UV}$ for the Lyman break
population of $U$-band dropouts at $z\approx 3$).  Subsequently, our
analysis of the UDF has been independently repeated by Yan \&
Windhorst (astro-ph/0407493) who also push to fainter magnitudes,
where the completeness corrections become significant. We are pleased
to report that the analysis of Yan \& Windhorst recovers almost all of
our original $i'$-band drop-out galaxies.

To avoid confusion over different naming conventions for the same
sources, we present here a matched list of ID numbers between our
first paper on the UDF $i'$-band drop-out galaxies (astro-ph/0403223),
and the recent preprint of Yan \& Windhorst (astro-ph/0407493), as
well as providing the corresponding identifications in the UDF v1.0
catalogue released by STScI\footnote{\tt ftp://archive.stsci.edu/pub/hlsp/udf/acs-wfc/h\_udf\_wfc\_V1\_z\_cat.txt}.
There is good agreement between the aperture magnitudes (with
an aperture correction appropriate for compact sources) reported in
our paper and the measurements of total
magnitudes (SExtractor `MAG\_AUTO') by Yan \& Windhorst from the same
images -- the typical difference is only 0.1\,mag in $z'$-band, and
the reported galaxy centroids typically agree to within
$0.03-0.06$\,arcsec ($1-2$\,pixels in the drizzled image).

Of the 53 brightest objects in the Yan \& Windhorst list (their IDs
\#1---48, including multiple sources \#1a,b, \#2a,b, \#5a,b,c, \#7a,b)
all are contained in our original list (plus \#58), with the exception
of just three objects.  Two of these (\#12 and \#41) lie in the noisy
edge region, outside the good quality central 11\,arcmin$^{2}$ within
which we restricted our original analysis. The third object, \#31, has
magnitudes $i'_{AB}=29.35$ and $z'_{AB}=28.21$ in our catalogue, so
its colour of $(i'-z')_{AB}=1.16$ from our photometry is slightly blue
of the $(i'-z')_{AB}>1.3$ cut adopted for $i'$-band drop-outs.  Hence
the catalogues agree at the 98\% level (one discrepant object out of
50).  We note that the two brightest $i'$-band drop-outs are probable
low-mass stars with $z'_{AB}\approx 25.3$: the Bunker et al.\ source
2140 (which also lies in the noisy edge region and appears as object
\#9 in the Yan \& Windhorst catalogue) was first identified as having
$i'$-drop colours but flagged as a probably star in Stanway, Bunker \&
McMahon (2003 MNRAS 342, 439, object SBM03\#5). The other probable
star is Bunker et al.\ source 11337 (object \#8 in the Yan \&
Windhorst catalogue). We note that our source 46574 (object \#3) has a
near neighbour clearly detected in $v$-band, and may be at low
redshift.

There are only 3 $i'$-band drop-outs which appeared in our original
catalogue and are not reproduced in the new Yan \& Windhorst list
(ignoring double sources): these are objects 49117, 14210 \& 42806 in
Bunker et al. (2004) We have inspected these again, and they appear
reasonable candidates, although there are nearby galaxies which are
not $i'$-drops which may influence the photometry.

\begin{table*}
\caption{$i$-band dropouts in the UDF. The two probable stars are above the line, all others are spatially resolved. 
}
\begin{tabular}{c|c|c|c|c|c|c}
\hline\hline
Bunker & Yan & STScI & RA \& Declination  &  $z'_{AB}$ (Bunker) & $z'_{AB}$ (Yan) & $\Delta z'$ (mag)  \\
et al.\ ID & ID & ID & J2000 (Bunker et al.)  & (0\farcs5-diam aper) & (total) & Bunker -- Yan  \\
\hline\hline
(2140)$^{\star}$ & \# 9    &    --- & 03 32 38.80  $-$27 49 53.6 &   25.22 $\pm$   0.02 &   25.43     &   -0.21      \\
 (11337)    &      \# 8    &    443 & 03 32 38.02  $-$27 49 08.4 &   25.29 $\pm$   0.02 &   25.41     &   -0.22      \\
\hline
20104$^1$   &      \#1a    &  2225 &  03 32 40.01  $-$27 48 15.0 &   25.35 $\pm$   0.02 &   25.25     &    0.10      \\        
42929$^2$   &      \#2a    &  8033 &  03 32 36.46  $-$27 46 41.4 &   26.56 $\pm$   0.03 &   26.49     &    0.07      \\        
41628       &      \#10    &  8961 &  03 32 34.09  $-$27 46 47.2 &   26.65 $\pm$   0.04 &   26.64     &    0.01      \\        
(46574)     &      \# 3    &  7730 &  03 32 38.28  $-$27 46 17.2 &   26.71 $\pm$   0.04 &   26.68     &    0.03      \\        
24019       &      \#11    &  3398 &  03 32 32.61  $-$27 47 54.0 &   26.80 $\pm$   0.04 &   26.68     &    0.12      \\        
52880       &      \#16    &  9857 &  03 32 39.07  $-$27 45 38.8 &   27.00 $\pm$   0.05 &   27.00     &    0.00      \\        
23516       &      \# 4    &  3325 &  03 32 34.55  $-$27 47 56.0 &   27.04 $\pm$   0.05 &   26.93     &    0.11      \\        
10188       &      \#15    &   322 &  03 32 41.18  $-$27 49 14.8 &   27.10 $\pm$   0.05 &   26.99     &    0.11      \\        
21422       &      \#20    &  2690 &  03 32 33.78  $-$27 48 07.6 &   27.23 $\pm$   0.05 &   27.35     &   -0.12      \\        
25578$^D$   &      \#13    &  ---  &  03 32 47.85  $-$27 47 46.4 &   27.30 $\pm$   0.06 &   26.97     &    0.33      \\        
25941       &      \# 6    &  4050 &  03 32 33.43  $-$27 47 44.9 &   27.32 $\pm$   0.06 &   27.23     &    0.09      \\        
26091$^D$   &      \#14    &  4110 &  03 32 41.57  $-$27 47 44.2 &   27.38 $\pm$   0.06 &   26.97     &    0.41      \\        
24458       &      \#23    &  3630 &  03 32 38.28  $-$27 47 51.3 &   27.51 $\pm$   0.07 &   27.50     &    0.01      \\        
21262       &      \#19    &  2624 &  03 32 31.30  $-$27 48 08.3 &   27.52 $\pm$   0.07 &   27.31     &    0.21      \\        
13494       &      \#18    & 30591 &  03 32 37.28  $-$27 48 54.6 &   27.56 $\pm$   0.07 &   27.25     &    0.31      \\        
24228       &      \#5b    &  3450 &  03 32 34.28  $-$27 47 52.3 &   27.63 $\pm$   0.07 &   27.17     &    0.46      \\  
16258       &      \#17    &  1400 &  03 32 36.45  $-$27 48 34.3 &   27.64 $\pm$   0.07 &   27.12     &    0.52      \\        
42414       &      \#22    &  9202 &  03 32 33.21  $-$27 46 43.3 &   27.65 $\pm$   0.07 &   27.39     &    0.26      \\        
27173       &      \#21    &  4377 &  03 32 29.46  $-$27 47 40.4 &   27.73 $\pm$   0.08 &   27.37     &    0.36      \\        
49117$^D$   &       ---    &  ---  &  03 32 38.96  $-$27 46 00.5 &   27.74 $\pm$   0.08 &             &              \\  
49701       &      \#28    & 36749 &  03 32 36.97  $-$27 45 57.6 &   27.78 $\pm$   0.08 &   27.72     &    0.06      \\        
24123       &      \#5a    &  ---  &  03 32 34.29  $-$27 47 52.8 &   27.82 $\pm$   0.08 &   26.97     &    0.85      \\        
27270       &      \#32    & 33003 &  03 32 35.06  $-$27 47 40.2 &   27.83 $\pm$   0.08 &   27.84     &   -0.01      \\        
23972       &      \#5c    &  3503 &  03 32 34.30  $-$27 47 53.6 &   27.84 $\pm$   0.09 &   27.76     &    0.08      \\  
14751       &      \#29    &  1086 &  03 32 40.91  $-$27 48 44.7 &   27.87 $\pm$   0.09 &   27.75     &    0.12      \\        
44154       &      \#7a    & 35945 &  03 32 37.46  $-$27 46 32.8 &   27.87 $\pm$   0.09 &   27.50     &    0.37      \\        
35084       &      \#26    & 34321 &  03 32 44.70  $-$27 47 11.6 &   27.92 $\pm$   0.09 &   27.65     &    0.27      \\        
42205       &      \#30    &  8904 &  03 32 33.55  $-$27 46 44.1 &   27.93 $\pm$   0.09 &   27.78     &    0.15      \\        
46503       &      \#36    &  7814 &  03 32 38.55  $-$27 46 17.5 &   27.94 $\pm$   0.09 &   27.97     &   -0.03      \\        
19953       &      \#1b    &  2225 &  03 32 40.04  $-$27 48 14.6 &   27.97 $\pm$   0.09 &   27.41     &    0.56      \\  
52086       &      \#34    & 36786 &  03 32 39.45  $-$27 45 43.4 &   27.97 $\pm$   0.09 &   27.89     &    0.08      \\        
44194       &      \#7b    & 35945 &  03 32 37.48  $-$27 46 32.5 &   28.01 $\pm$   0.10 &   27.78     &    0.23      \\  
21111$^D$   &      \#33    &  2631 &  03 32 42.60  $-$27 48 08.9 &   28.02 $\pm$   0.10 &   27.86     &    0.16      \\        
46223$^D$   &      \#24    & 35506 &  03 32 39.87  $-$27 46 19.1 &   28.03 $\pm$   0.10 &   27.61     &    0.42      \\        
22138       &      \#40    & 32007 &  03 32 42.80  $-$27 48 03.2 &   28.03 $\pm$   0.10 &   28.06     &   -0.03      \\        
14210       &       ---    &  978  &  03 32 35.82  $-$27 48 48.9 &   28.08 $\pm$   0.10 &             &              \\  
45467       &      \#38    & 35596 &  03 32 43.02  $-$27 46 23.7 &   28.08 $\pm$   0.10 &   28.00     &    0.08      \\        
12988$^D$   &      \#25    & 30534 &  03 32 38.49  $-$27 48 57.8 &   28.11 $\pm$   0.11 &   27.63     &    0.48      \\        
30359       &       ---    & 33527 &  03 32 30.14  $-$27 47 28.4 &   28.13 $\pm$   0.11 &             &              \\  
11370       &      \#45    &   482 &  03 32 40.06  $-$27 49 07.5 &   28.13 $\pm$   0.11 &   28.20     &   -0.07      \\        
24733       &      \#27    & 32521 &  03 32 36.62  $-$27 47 50.0 &   28.15 $\pm$   0.11 &   27.65     &    0.50      \\        
37612       &      \#37    & 34715 &  03 32 32.36  $-$27 47 02.8 &   28.18 $\pm$   0.11 &   27.99     &    0.19      \\        
41918       &      \#46    &  7829 &  03 32 44.70  $-$27 46 45.5 &   28.18 $\pm$   0.11 &   28.27     &   -0.09      \\        
21530       &      \#42    & 31874 &  03 32 35.08  $-$27 48 06.8 &   28.21 $\pm$   0.12 &   28.13     &    0.08      \\        
42806$^2$   &      \#2b    &  8033 &  03 32 36.49  $-$27 46 41.4 &   28.21 $\pm$   0.12 &   27.76     &    0.45      \\  
27032$^D$   &      \#21    &  4377 &  03 32 29.45  $-$27 47 40.6 &   28.22 $\pm$   0.12 &   27.37     &    0.85      \\        
52891       &      \#43    & 36697 &  03 32 37.23  $-$27 45 38.4 &   28.25 $\pm$   0.12 &   28.14     &    0.11      \\        
17908       &      \#35    &  1834 &  03 32 34.00  $-$27 48 25.0 &   28.25 $\pm$   0.12 &   27.94     &    0.31      \\        
48989$^D$   &      \#39    & 36570 &  03 32 41.43  $-$27 46 01.2 &   28.26 $\pm$   0.12 &   28.00     &    0.26      \\        
17487       &      \#47    &  ---  &  03 32 44.14  $-$27 48 27.1 &   28.30 $\pm$   0.12 &   28.30     &    0.00      \\        
18001       &      \#48    & 31309 &  03 32 34.14  $-$27 48 24.4 &   28.40 $\pm$   0.13 &   28.38     &    0.02      \\        
35271       &      \#44    &  6325 &  03 32 38.79  $-$27 47 10.9 &   28.44 $\pm$   0.14 &   28.16     &    0.28      \\        
22832       &      \#58    &   --- &  03 32 39.40  $-$27 47 59.4 &   28.50 $\pm$   0.15 &   28.58     &   -0.08      \\        
\hline
\end{tabular}
\label{tab}
\ \\
$^{D}$\,Double.\ $^{\star}$\,star SBM03\# 5, on image edge.\
$^{1}$\, spectroscopic $z=5.83$ (SBM03\#1 in Stanway et al.\ 2004 ApJ 607, 704; SiD002 in Dickinson et al.\ 2004 ApJL 600, 99; 
GLARE\,1042 in Stanway et al.\ 2004 ApJL 604, 13).\ 
$^{2}$\,SiD025 (Dickinson et al.\ 2004) 42929 \& fainter 42806.\
\end{table*}

\end{document}